\documentclass[pra,aps,preprint]{revtex4}
\usepackage{amsmath}
\usepackage{graphicx,pstricks,pst-coil,pst-plot,pst-slpe}

\input{tcilatex}

\begin{document}

\title{An quantum approach of measurement based on the Zurek's triple model }
\author{P.Zhang$^{1}$,X.F.Liu$^{1,2},$C.P.Sun$^{1,a,b}$}
\address{1.Institute of Theoretical Physics,Academia Sinica,Beijing,
China\\
2.Department of Mathematics,Peking
University,Beijing,China\medskip}

\begin{abstract}
In a close form without referring the time-dependent Hamiltonian
to the total system, a consistent approach for quantum measurement
is proposed based on Zurek's triple model of quantum decoherence
[W.Zurek, Phys. Rev. D 24, 1516 (1981)]. An exactly-solvable model
based on the intracavity system is dealt with in details to
demonstrate the central idea in our approach: by peeling off one
collective variable of the measuring apparatus from its many
degrees of freedom, as the pointer of the apparatus, the
collective variable de-couples with the internal environment
formed by the effective internal variables, but still interacts
with the measured system to form a triple entanglement among the
measured system, the pointer and the internal environment. As
another mechanism to cause decoherence , the uncertainty of
relative phase and its many-particle amplification can be summed
up to an ideal entanglement or an Shmidt decomposition with
respect to the preferred basis.

\end{abstract}

\maketitle

\address{$^1$Institute of Theoretical Physics,Academia Sinica,Beijing,
100080,China\\
$^2$Department of Mathematics, Peking University, Beijing,
100871,China }

\section{Introduction}

von Neumann's quantum theory of measurement emphasizes that, after a
measurement, the emergence of classicality of a quantum system $S$ from
quantum dynamics is due to a perfect correlation between this system and its
measuring apparatus $A$ described by quantum mechanics [1]. But W.H. Zurek
argued that this theory does not thoroughly solve the core problem in
quantum measurement [2]. His argument is, the interaction between the
quantum system and the apparatus can only produce a quantum entanglement
like the EPR state with quantum uncertainty [3], rather than a classical
correlation described by a statistical operator with classical probability
distribution $|c_{s}|^{2}$- the density matrix $\rho _{c}=$ $%
\sum_{s}|c_{s}|^{2}\left\vert s\right\rangle \langle s|\otimes
|p_{s}\rangle \langle p_{s}|$ where $\left\vert s\right\rangle $
and$\left\vert p_{s}\right\rangle $ are orthonormal basis vectors
of the system to be measured and the pointer state of the
apparatus respectively. To go beyond von Neumann's theory, Zurek
proposed an elegant \textquotedblright triple
model\textquotedblright\ for quantum measurement twenty years ago.
In his theory, besides the quantum system and the apparatus, an
environment $E$ must be introduced as a necessary element to
generate the triple entanglement

\begin{equation}
\left| \Phi _{Tri}\right\rangle =\sum_sc_s\left| s\right\rangle \otimes
\left| p_s\right\rangle \otimes \left| e_s\right\rangle
\end{equation}
through the coupling of the apparatus to the environment. It is obvious that
the classical mixture state $\rho _c$ of correlation can be obtained by
ignoring ( mathematically ''tracing over '') the environment states.

Zurek's triple model in principle overcomes the key difficulty in quantum
measurement theory, but it still needs microscopic refinement in terms of
quantum dynamics and there remain details to be filled in. Actually, just as
Zurek points out, to implement such triple entanglement as dynamic Schmidt
decomposition, the interactions among the triple parts should be time
dependent. To be more precise, two steps are required to finish the
measurement: turning on the interaction $H_{SA}$ between $S$ and $A$ at the
instant $t=0$ and turning on the interaction $H_{AE}$ between $A$ and $E$ at
another instant $t=t_{m}$. The process can be represented as follows:
\begin{equation}
\sum_{s}c_{s}\left\vert s\right\rangle \otimes \left\vert p\right\rangle
\otimes \left\vert e\right\rangle _{t=0}\overset{H_{SA}}{\rightarrow }\left(
\sum_{s}c_{s}\left\vert s\right\rangle \otimes \left\vert p_{s}\right\rangle
\right) \otimes \left\vert e\right\rangle _{t=t_{m}}\overset{H_{AE}}{%
\rightarrow }\left\vert \Phi _{Tri}\right\rangle
\end{equation}

However, the time dependence of Hamiltonian means that there exists another
extra system governing the ''universe'' formed by the triple system. So the
quantum dynamic theory describing the measurement is not in a close form.
Moreover, to realize a real measurement process, one should switch the
couplings at certain exact instants. In practice, it is difficult to exactly
control the interaction between $A$ and $E$ so that it occurs only after the
correlation between $A$ and $S$ has just been established. Another point we
wish to mention is that according to Zurek's model, to produce an ideal
triple entanglement such that $\{\left| s\right\rangle \},\{\left|
p_s\right\rangle \}$and $\{\left| e_s\right\rangle \}$form three orthonormal
sets, it is even required that there is no interaction between $S$ and $E$
as described in Eq.(1); otherwise the Schmidt decomposition structure of
Eq.(1) would be destroyed.

Most recently, we study the phenomenon of quantum decoherence of a
macroscopic object along a different direction: we investigate the adiabatic
quantum entanglement [4] between its collective states (such as that of the
center-of-mass (C.M)) and its inner states. It is shown that the adiabatic
wave function of a macroscopic object can be written as an entangled state
with correlation between adiabatic inner states and quasi-classical motion
configurations of the C.M. Since the adiabatic inner states are factorized
with respect to the composing parts of the macroscopic object [5], this
adiabatic separation can induce quantum decoherence. This observation thus
provides us with a possible solution to the Schroedinger cat paradox. This
approach to quantum decoherence only concerns a double system rather than a
triple one, so it does not solve the quantum measurement problem completely.
Rather, it provides a novel example for von Neumann's quantum measurement
theory, which does not produce a classical correlation.

In this paper, integrating the above mentioned results with Zurek's triple
model, we present a consistent quantum -mechanical approach for measurement
process in a close form with time-independent total Hamiltonian. In this
alternative, the measuring apparatus is taken as a macroscopic object with
effective inner variables and a pointer variable. The novel point in our
treatment lies in the complete separation of the pointer variable from the
effective inner variables of the macroscopic apparatus $A$. With this
separation there is no coupling between the pointer variable and the inner
variables of $A,$but the effective interaction of the pointer with $S$ is
induced by that of the original variables of $A$ in an adequate way. Just
for this reason, the triple entanglement [2] can be dynamically generated
without the time-dependent control. To sketch our basic idea, we start with
an exactly solvable model in the intracavity dynamics. Using this example we
also show that the back action of the inner environment on the system plus
pointer implied by Heisenberg's position-momentum uncertainty relation will
disturb the phases of states between the system and pointer and then
decohers the quantum entanglement system plus pointer, which is formed
dynamically just before measurement.

\section{Outline of Our Approach for Quantum Measurement Based on Zurek's
Theory}

The quantum theory of measurement based on von Neumann's theory usually
treats the measuring process as a quantum mechanical evolution by
considering the measuring apparatus $A$ as a proper quantum system . This is
just in contrast to the Copenhagen interpretation with the hypothesis of
classicality on the part of the apparatus. According to the theory of
Copenhagen school the apparatus should behave classically so that the
experimental outcome of measurement can be recorded in the classical way.
Zurek's theory does not stress the classicality of apparatus directly since
the meaning of classicality of apparatus is not clear without association
with the measured system $S$. The important discovery by Zurek is the
decoherence of the quantum entanglement between the measuring apparatus and
the measured system induced by an external or inner environment $E$ . Led by
Zurek's observation one may imagine that, it is the direct interaction of
environment with the pointer of apparatus that leads to the classicality of
apparatus. However, it is not true !

In the following , we can show that, to decoher the quantum entanglement
between $A$ and $S$, there only need two proper couplings of the measured
system to the pointer and to the environment ,and the interaction between $A$
and $E$ is not necessary. The requirement of ''no interaction between $A$
and $E$ '' will result in a time -independent re-formulation of Zurek's
triple theory . In our quantum approach of measurement by generalizing
Zurek's triple theory, we consider $E$ as the collection of the internal
relative degrees of freedom. Then, the pointer (also denoted by $A)$ of the
apparatus can be defined as the collective (or macroscopic) degree of
freedom of the apparatus, e.g., the coordinate of the center of mass (C.M).

In general we write a time-independent total Hamiltonian as
\begin{equation}
H=H_s+H_a+H_e+V_{sa}+V_{se}
\end{equation}
Here, $H_s=H_s(q_s),$ $H_a=$ $H_a(q_a)$ and $H_e=H_e(q_e)$ are respectively
the free Hamiltonian for $S,A$ and $E$.; $q_s,q_a$ and $q_e$ roughly stand
for the system variable, the pointer variable and the environment variable
correspondingly; $V_{sa}=$ $V_{sa}(q_a,q_s)$ describes the interaction
between $S$ and $A$ while $V_{se}=V_{se}(q_e,q_s)$ describes that between $S$
and $E$.

To gain a close form for quantum measurement based on Zurek's triple theory,
it is important that no interaction exists between the pointer $A$ and the
''inner'' environmental '' $E$ (in Fig.1). Only by assuming that the system
satisfies the following double non-demolition condition:
\begin{equation}
\lbrack H_s,V_{sa}]=0,[H_s,V_{se}]=0
\end{equation}
the evolution operator for the total system can be written as

\begin{equation}
U(t)=\sum_ne^{-iE_nt}\left| n\right\rangle _{ss}\left\langle n\right|
U_{an}(t)\otimes U_{en}(t)
\end{equation}
Here, \{$\left| n\right\rangle _s$\} is an eigenvector of $H_s$
corresponding to the eigenvalue $E_n$ ,
\begin{equation*}
U_{an}=_s\langle n|\exp [H_a(q_a)+V_{sa}(q_a,q_s)]\left| n\right\rangle _s
\end{equation*}
and
\begin{equation*}
U_{en}=_s\langle n|\exp [-i(H_e(q_e)+V_{se}(q_e,q_s))]\left| n\right\rangle
_s
\end{equation*}
are the effective evolution operators. They describe the feedbacks of the
measured system on the pointer of apparatus and the environment respectively
when $S$ is just in its eigen-state $\left| n\right\rangle _s.$

%
\begin{figure}[h]
\begin{center}
\includegraphics[width=10cm,height=7cm]{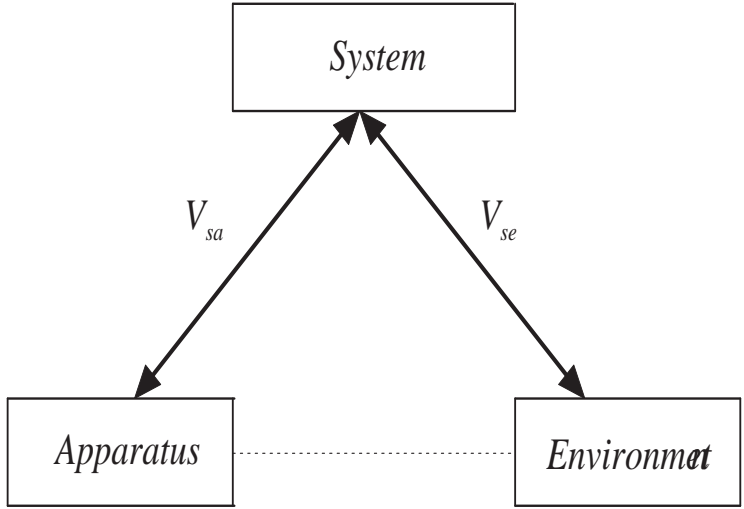}
\end{center}
\caption{The Effective interactions among the system $S$, the pointer $A$
and the inner environment $E$}
\end{figure}
It is worthy to point out that, in the present approach for entanglement,
the energy of the measured system is conserved while the quantum coherence
is destroyed.

This kind of unitary evolution operator $U(t)$ can establish a non-separable
correlation among the system, the pointer and the environment. Namely, if
the initial state $\left| \Psi \right\rangle _{initial}=\left|
s\right\rangle \otimes \left| a\right\rangle \otimes \left| e\right\rangle $
of the total system is of a factorized form with the system state $\left|
s\right\rangle $ $=\sum_nc_n\left| n\right\rangle _s$ , the pointer state $%
\left| a\right\rangle $ and the environment state $\left| e\right\rangle $
,then the final state of the total system will be

\begin{equation}
\left| \Psi \right\rangle _{final}=U(t)\left| \Psi \right\rangle
_{initial}=\sum_nc_ne^{-iE_nt}\left| n\right\rangle _s\otimes \left|
a_n\right\rangle \otimes \left| e_n\right\rangle
\end{equation}
Here, $\left| a_n\right\rangle =U_{an}\left| a\right\rangle $ and $\left|
e_n\right\rangle =U_{en}\left| e\right\rangle $ are the final states of the
pointer and the environment entangling with the system states $\left|
n\right\rangle _s.$ Thus, a triple correlation among the measured system,
the pointer and the environment is established. Obviously , the reduced
density matrix for the composite subsystem formed by the system plus the
pointer is
\begin{eqnarray}
\rho &=&\sum_n\left| c_n\right| ^2\left| n\right\rangle _{ss}\left\langle
n\right| \otimes \left| a_n\right\rangle \left\langle a_n\right|  \notag \\
&&+\sum_{m\neq n}c_m^{*}c_n\left| n\right\rangle _{ss}\left\langle m\right|
\otimes \left| a_n\right\rangle \left\langle a_m\right| \langle e_m\mid
e_n\rangle
\end{eqnarray}
The off-diagonal terms on the r.h.s of this equation is responsible for the
interference pattern. It is easy to see that the interference fringe
completely vanishes when the states of the inner part $E$ are orthogonal to
one another. In this situation, an ideal Zurek$^{\prime }$s classical
correlation
\begin{equation}
\rho =\sum_n\left| c_n\right| ^2\left| n\right\rangle _{ss}\left\langle
n\right| \otimes \left| a_n\right\rangle \left\langle a_n\right|
\end{equation}
results from the ideal entanglement with the correlated components $%
|e_n\rangle $ orthogonal to one another.

The above Zurek$^{\prime }$s classical correlation [2] just describes the
fact like weather forecast impersonally predicting whether it rains or not
tomorrow. The equation (7) deterministically tells us the classical
correlation that the system is in $|n\rangle $ when the pointer is just in $%
\left| a_n\right\rangle $ with probability $|c_n|^2.$ This is unlike the
quantum entanglement $|s\rangle =\sum_nc_n\left| n\right\rangle \otimes
\left| a_n\right\rangle $ that not only indicates the correlation between $%
\left| n\right\rangle $ and $\left| a_n\right\rangle $ , but also
simultaneously tells us the correlation with probability $p_{n\text{ }%
}=\sum_{n^{^{\prime }}}|s_{n^{^{\prime }}n}^{-1}c_{n^{^{\prime }}}|^2$
between any superposition state $\left| s_n\right\rangle =\sum_{n^{\prime
}}s_{nn^{^{\prime }}}\left| n^{^{\prime }}\right\rangle $ of $S$ and the
corresponding one
\begin{equation*}
\left| t_n\right\rangle =\sqrt{\frac 1{p_n}}\sum_{n^{^{\prime
}}}s_{n^{^{\prime }}n}^{-1}c_{n^{^{\prime }}}\left| a_{n^{^{\prime
}}}\right\rangle
\end{equation*}
of $A$ . This is because $|s\rangle $ can also be re-expressed as
\begin{equation}
|s\rangle =\sum_np_n\left| s_n\right\rangle \otimes \left| t_n\right\rangle .
\end{equation}
In fact, the classical correlation does not say anything about the
correlation of different pairs $\left| s_n\right\rangle $ and $\left|
t_n\right\rangle $ but for the original pair $\left| n\right\rangle $ and $%
\left| a_n\right\rangle $, and its prediction is independent of what to be
measured. On the contrary, what the quantum entanglement tells us depends on
what we measure according to EPR argument [3]. With this understanding, it
can be said that quantum measurement is implemented completely when quantum
decoherence happens to result in the so called classical correlation.

The above argument shows that the vanishing overlap of the final states of
the inner environment is necessary to obtain the classical correlation after
measurement.We define this overlap $F_{m,n}=\langle e_m|e_n\rangle $ as
decoherence factor. Now an immediately-following question is in what case
the decoherence factor becomes zero. Our previous works on quantum
measurement theory [5] showed that an ideal entanglement appears in the
macroscopic limit that the number $N$ of particles making up the detector
approaches infinity. In the present case, we assume there are $N$ degrees of
freedom in the apparatus. We can peel off one (or more ) collective variable
as the pointer of the apparatus, and the inner environment is formed by $N$
relative internal variables $q_k(k=1,2,..N)$. We can imagine that there are $%
N$ blocks constituting the inner environment, so we may write $%
H_e(q_e)=\sum_kH_e^{(k)}(q_k)$ and $V_{se}(q_e,q_s)=%
\sum_kV_{se}^{(k)}(q_k,q_s)$ in sum forms. If all $%
V_{se}^{(k)}(q_k,q_s)(k=1,2,..N)$ commute with one another, we can factorize
the effective evolution operator $U_{en}$:
\begin{equation*}
U_{en}=\prod_{j=1}^NU_{en}^{[j]}
\end{equation*}
When the measured system is initially prepared in $\left| n\right\rangle $
and the environment in a factorized state $|e\rangle
=\prod_{j=1}^N|e^{[j]}\rangle ,$ the environment will obey a factorized
evolution
\begin{equation}
|e\rangle \rightarrow |e_n^{[j]}\rangle \equiv
\prod_{j=1}^N|e_n^{[j]}\rangle =\prod_{j=1}^NU_{en}^{[j]}|e^{[j]}\rangle q
\end{equation}
entangling with the system state $\left| n\right\rangle .$ It results in
the\ factorization structure [5] of decoherence factor

\begin{equation}
F_{m,n}=\prod_{j=1}^N\ \langle e_m^{[j]}|e_n^{[j]}\rangle .
\end{equation}
Since each factor $\langle e_m^{[j]}|e_n^{[j]}\rangle $ in $F_{m,n}$ has a
norm less than unity, the product of infinite such factors may approach
zero. This investigation was developed based on the Hepp-Coleman model and
its generalizations [12,13] and was applied to analyzing the universality of
the influence of environment on quantum computing process [14].

\section{ Pointer Model for Quantum Measurement in Intracavity System}

In this section and the subsequent sections, we will use an exactly-solvable
model based on the over-simplified intracavity system to demonstrate our
central ideas .

Consider a cavity with two end mirrors (as in Fig.2), one of which is fixed
while the other is treated as a macroscopic object consisting of $N$
particles of mass $m_i$ with position coordinate $x_i$ and momentum
coordinate $p_i$($i=1,2,..N$). The radiation pressure of the cavity field on
the moving mirror is proportional to the intracavity photon density. Let $%
a^{\dagger }$ and $a$ be the creation and annihilation operators of the
cavity with a single mode of frequency $\omega $. The cavity-mirror coupling
is described by an interaction Hamiltonian $H_I=-\sum_i^Ng_ix_i$ $a^{\dagger
}$ $a$ where $g_i$ is coupling constant depending on the electric dipole.
%
\begin{figure}[h]
\begin{center}
\includegraphics[width=10cm,height=7cm]{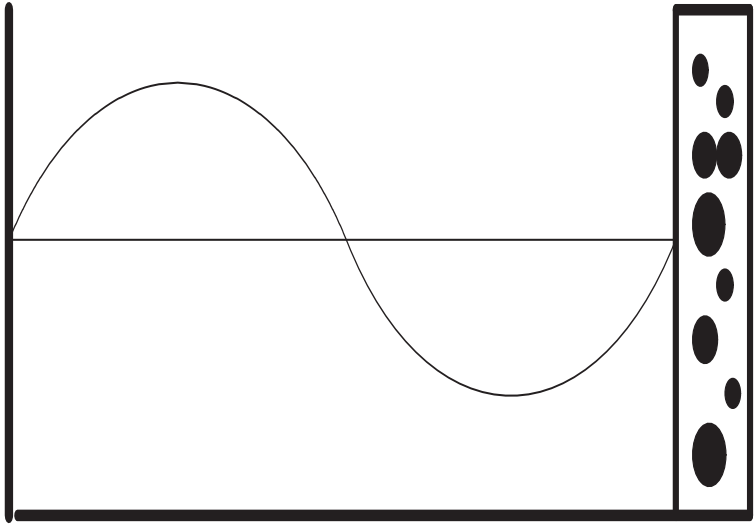}
\end{center}
\caption{\textit{A cavity with a moving end mirror B and a fixed one A. The
moving one is treated as an macroscopic object consisting of $N$ free
particles }}
\end{figure}
In this situation we describe the cavity field dynamics with the free
Hamiltonian $H_s=\omega _0a^{\dagger }$ $a$. This cavity field -mirror
coupling system can also be used to detect the photon number in the cavity
by the motion of the mirror. Obviously, the total Hamiltonian governing the
motion of the mirror is
\begin{equation}
H=\omega _0a^{+}a+\sum_{i=1}^N\frac{p_i^2}{2m_i}-a^{+}a\sum_i^Ng_ix_i
\end{equation}
By taking the moving mirror as a whole, this intracavity model is associated
with the interferometric detection of the gravitational wave [6-8]. This
system has already been studied quite extensively by many authors [8], under
the assumption that the moving mirror is a macrocosm without internal
structure. Like the discussions in these previous references [6-8], we also
assume the oscillation of the movable mirror is so small that the the
frequency of single cavity mode does not undergo a considerable change due
to the displacement of the mirror.

The above described physical system can be viewed as a quantum measurement
system, in which the cavity field monitored by the macroscopic moving mirror
is to be measured . Since $[H_s,H_I]=0$ and $[H_s,\sum_i\frac{p_i^2}{2m_i}%
]=0 $ , the moving mirror has no influence on the cavity field . On the
other hand, different eigen-states $|n\rangle $ of $H_s$ can imprint on the
moving mirror differently. Indeed, the interaction term $-n\sum_i^Ng_ix_i$
implies that different forces will act on the moving mirror when the cavity
is prepared in different number states $|n\rangle .$ We notice that this is
a typical characteristic of non-demolition measurement [9].

We take the CM position $x=M^{-1}\sum_i^Nm_ix_i$ to be the pointer of the
apparatus - the moving mirror and the relative coordinates $\xi _j=x_j-x$ $%
(j=1,2,...,N-1)$ to be the internal variables. Here, $M=\sum_{i=1}^Nm_i$ is
the total mass of the moving mirror. We denote the conjugate momenta of $x$
and $\xi _j$ by $p_x$ and $p_{\xi j}$. Then, we obtain an interesting
realization of Zurek' s triple model with the time-independent Hamiltonian
\begin{equation}
H=H_s+H_a+H_e+V_{sa}+V_{se}
\end{equation}
where $H_a=\frac{p_x^2}{2M}$ and
\begin{equation}
H_e=\frac 12\sum_{i,j=1}^{N-1}\tau ^{-1}{}_{ij}p_{\xi i}p_{\xi j}
\end{equation}
are the free Hamiltonians for the pointer part and the internal environment
respectively,
\begin{equation}
V_{sa}=-Ga^{+}ax
\end{equation}
is the effective interaction describing the coupling of the system to the
pointer and
\begin{equation}
V_{se}=-\sum_{j=1}^{N-1}G_ia^{+}a\xi _j
\end{equation}
describes an interaction between the system and the internal environment.
Here,
\begin{equation}
G_i=(g_i-\frac{m_i}{m_N}g_N),G=\sum_{i=1}^{N-1}g_i
\end{equation}
are the effective coupling constants, and the mass matrix $\tau $ is defined
by the matrix elements
\begin{equation}
\tau _{ij}=m_i\delta _{ij}+\frac{m_im_j}m_N
\end{equation}
This expression of $\tau $ is obtained by substituting the individual
laboratory coordinate
\begin{equation}
x_N=\frac 1m_N[x-\frac 1M\sum_i^{N-1}m_i(\xi _j+x)]
\end{equation}
into the interaction term $-a^{+}a\sum_i^Ng_ix_i$ of the total system
Hamiltonian.

In this model the couplings of the system to the pointer and to the
environment can be turned - on or turned -off simultaneously for $V_{sa}$
and $V_{se}$ are proportional to the same coupling constant. Physically this
is reasonable as the pointer variable is a reflection of the collective
average effect of the internal degrees of freedom of the macroscopic
apparatus. Another remarkable character of the above model is the absence of
coupling between the pointer variable and the inner variables of $A.$
Physically this is just what is required of an ideal measuring apparatus,
whose effective inner motion should not directly affect the reading of the
pointer. In fact this property guarantees that a triple entanglement will
form dynamically from a factorized initial state. This just realizes our
central ideas in section 2: the de-coupling of pointer variable with the
inner ones in the apparatus and the factorization of the inner environment
with respect to each inner variable.

\section{Exact Solution with Factorization for Ideal Entanglement}

Now let us consider the exact solution to the dynamic evolution problem of
the intracavity model introduced in section II. To this end we invoke a
canonical transformation
\begin{equation}
\eta _i=\sum_{j=1}^{N-1}U_{ij}\xi _j,i,j=1,...,N-1
\end{equation}
from the inner variables $\xi _j$ to a new set of canonical variables $\eta
_i.$ Here , $U$ is an ($N-1)$ by ($N-1)$ matrix diagonalizing the mass
matrix $\tau $ , i.e. $U\tau U^T$ is diagonal. Denote the conjugate momenta
of $x$ and $\eta _i$ by $p_x$and $p_i$ respectively. The inner environment
is described by the new internal variable $\{\eta _i\}$. Then, the total
Hamiltonian can be re-expressed as
\begin{equation}
H=\frac{p_x^2}{2M}-Ga^{+}ax+\sum_{i=1}^{N-1}(\frac 1{2m_i^{^{\prime
}}}p_i^2-f_ia^{+}a\eta _i)+\omega _0a^{+}a
\end{equation}
where the eigenvalues $m_i^{^{\prime }}$ of matrix $\tau $ are the effective
mass with respect to the new coordinates $\{\eta _j\};$ $f_i=%
\sum_jG_jU_{ji}^T$ represent the strengths of forces on each inner
coordinate $\eta _i$ by one photon of cavity field. Obviously, the inner
motion of the moving mirror is factorizable since the effective Hamiltonian $%
H_{se}=\sum H_{se}^j$ is only a simple sum of the single component
\begin{equation}
H_{se}^j(a^{+}a)=\frac 1{2m_i^{^{\prime }}}p_i^2-f_ia^{+}a\eta _i
\end{equation}
This direct sum structure results in the factorization of the effective
evolution matrix defined by
\begin{eqnarray}
U_{en}(t) &=&\langle n|U(t)|n\rangle =\prod_{i=1}^{N-1}U_{en}^i(t):  \notag
\\
U_{en}^i(t) &=&\exp [-i(\frac{p_i^2}{2m_i^{^{\prime }}}-nf_i\eta _i)t] \\
&=&e^{-i\frac{tp_i^2}{2m_i^{^{\prime }}}}e^{i\frac{nf_it^2p_i}{%
2m_i^{^{\prime }}}}\exp [inf_it\eta _i-i\frac{n^2f_i^2t^3}{6m_i^{^{\prime }}}%
]
\end{eqnarray}
where $|n\rangle $ is the Fock state and $U(t)=\exp (-iHt)$ is the evolution
matrix of the total system. Here, we have used the Wei-Norman algebraic
method [10] to re-write the single particle evolution matrix .

Without loss of generality, we assume each inner component of the moving
mirror to be described by the same Gaussian wave packet $\mid i\rangle $ of
width $a,$ i.e.
\begin{equation}
\langle \eta _i\mid i\rangle =\left( \frac 1{2\pi a^2}\right) ^{\frac 14}e^{-%
\frac{\eta _i^2}{4a^2}}
\end{equation}
This is a physically reasonable preparation of initial state. This is
because the initial state of the inner $\Phi (0,\{\eta
_j\})=\prod_{i=1}^{N-1}\langle \eta _i|i\rangle $ also defines the
factorized Gaussian wave packet $\Phi (0,\{\xi
_i\})=\prod_{i=1}^{N-1}\langle \xi _i|i\rangle $ in $\xi $- representation
since $\xi _i^2$ is a canonical invariant, i.e., $\sum_i\xi _i^2=\sum_i\eta
_i^2.$ A Gaussian wave packet $\langle \xi _i|i\rangle \simeq \exp [-\frac{%
(x_i-x)^2}{4a^2}]$ implies basically that the particles composing the moving
mirror are almost peaked on the position of C.M. We also assume that the C.M
of the moving mirror is just in the position eigenstate $\left|
X\right\rangle $ and the cavity is initially in a coherent state $|\alpha
\rangle =\sum_nc_n\left| n\right\rangle $ where $c_n=e^{-\frac 12|\alpha |^2}%
\frac{\alpha ^n}{n!}$. In this case using the effective evolution matrix of
the pointer
\begin{equation}
U_{nx}=\exp [-i(\frac{p_x^2}{2M}-Gnx])t]
\end{equation}
with respect to the cavity Fock state $\left| n\right\rangle $ , we
explicitly obtain the triple entanglement
\begin{equation}
\left| \Psi (t)\right\rangle =\sum_nc_n(t)\left| n\right\rangle \otimes
\left| x_n\right\rangle \otimes \left| e_n\right\rangle .
\end{equation}
at any instance $t\neq 0$ . Here ,
\begin{equation*}
c_n(t)=c_ne^{-i(n+1/2)\omega _0t},\left| x_n\right\rangle =U_{nx}\left|
x\right\rangle
\end{equation*}
and
\begin{equation*}
\left| e_n\right\rangle =U_{en}(t)|\Phi (0)\rangle
=\prod_{j=1}^NU_{en}^j(t)\mid j\rangle
\end{equation*}
We can also calculate the decoherence factor with factorization
\begin{eqnarray*}
F_{mn}(t) &=&\langle e_n\left| e_m\right\rangle =\prod_{j=1}^N\langle
e_n^{[j]}\left| e_m^{[j]}\right\rangle \\
&\equiv &\prod_{j=1}^N\ \langle j|U_{en}^{j\dagger }(t)U_{em}^j(t)j\rangle
\end{eqnarray*}
to give the decaying norm
\begin{equation}
f_{mn}(t)=\mid F_{mn}(t)\mid =\exp [-(n-m)^2\sum_{j=1}^{N-1}f_i^2(\frac{t^4}{%
32m_i^{^{\prime 2}}}+\frac{a^2t^2}2)]
\end{equation}
In Fig.2, the decaying behavior of decoherence factor is demonstrated for
different $N$. It is seen that a decoherence process indeed happens as $%
t\rightarrow \infty ,$ but it does not obey the simple exponential law $%
e^{-\gamma t}.$ In a long time scale, the temporal behavior of decoherence
is described by
\begin{equation*}
F(t)\approx \exp \left[ -(m-n)^2\Gamma t^4\right]
\end{equation*}
with $\Gamma $ $=\sum_{j=1}^{N-1}\frac{f_i^2}{32m_i^{\prime 2}a^2}.$ If we
define the characteristic time $\tau _d$ of the decoherence process by $%
F(\tau _d)=e^{-1}$ , then
\begin{equation}
\tau _d=[(m-n)^2\Gamma ]^{1/4}
\end{equation}
This shows that the long time behavior of decoherence depends directly on
interaction.

%
\begin{figure}[h]
\begin{center}
\includegraphics[width=10cm,height=7cm]{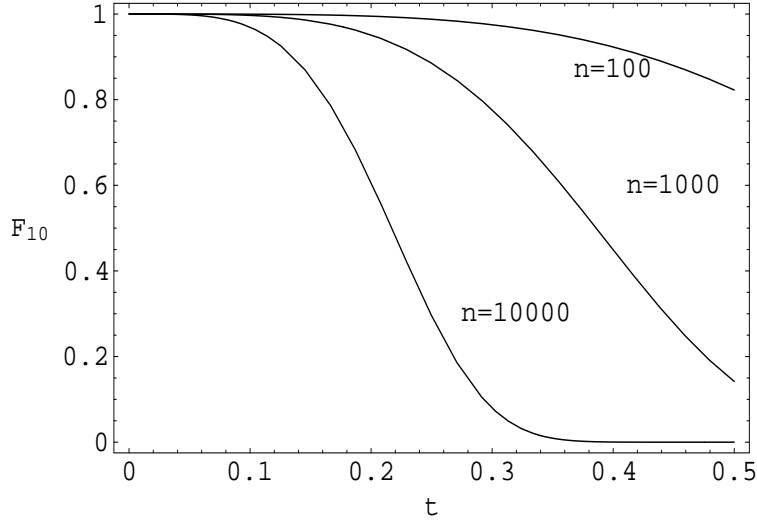}
\end{center}
\caption{The decoherence factor $F_{10}$ plotted as a function of time and
for various values of the particle number $N$. We have taken the mass of a
particle $m=10^{-6}$kg, the wave width of the Gaussian wave packet as the
length of the cavity $a=10^{-5}$m and the force strength on each inner
coordinate $f=10^{-14}$kgms${^{-2}}$. The real particle number $N=10^6n$. }
\end{figure}
In the macroscopic limit $N\rightarrow \infty $ or for the long time
evolution $t\rightarrow \infty ,$ the vanishing decoherence factors $\langle
e_n\left| e_m\right\rangle $ leads to an ideal triple entanglement (1) with
invariant probability distribution $p_n=|c_n(t)|^2=|c_n|^2$ .

Finally we need to show the measuring process we modelled above is ideal
since the pointer states entangling with each system state are orthogonal to
one another, namely , $\langle x_n\left| x_m\right\rangle $ $\sim \delta
_{n,m}$ for $m\neq n$ .In fact, in the coordinate-representation, the
pointer state $\left| x_n\right\rangle $ can be calculated explicitly as
\begin{equation}
\langle x|x_n\rangle =\sqrt{\frac{-iM}{2\pi t}}\exp [i\left\{ \frac{\left[
nGt^2+2M\left( x-X\right) \right] ^2}{8Mt}+\left( nGtX-\frac{n^2G^2t^3}{6M}%
\right) \right\} ]
\end{equation}
This means that the width of each wave packet $\langle x|x_n(t)\rangle $ is
zero and then the overlaps
\begin{equation}
|O_{n,m}|=|\langle x_n(t)|x_m(t)\rangle |=\frac{2M}{Gt^2}\delta \left(
m-n\right)
\end{equation}
of wave packets vanish for $m\neq n$ . This indicates an ideal
classical correlation between the measured system and the pointer
of apparatus in our intracavity model. Therefore, our triple model
for quantum measurement leads to an ideal quantum measurement in a
purely quantum dynamical way with neither the introduction of the
hypothesis of classicality for the apparatus, nor the artificial
control of interactions.

As shown above, an ideal entanglement of the double system (formed by the
measured system plus pointer of apparatus) with an ( inner ) environment is
a necessary element to force the apparatus to behave so classically that a
de-entanglement process occurs in the double system. Another necessary
element to implement quantum measurement is the ideal entanglement between
the measured system and the pointer of apparatus. The work of this mechanism
depends on the choice of the form of initial state of the pointer. For a
general initial state of the pointer, it is difficult to obtain an explicit
condition under which $\langle $ $x_n\left| x_m\right\rangle $ becomes zero.
The solution to this problem concerns the consideration of the classical
limit of the motion of the pointer as a large system. In general, for
certain particular states, the classical limit of the expectation value of
an observable should recover its classical value form. Such quasi-classical
states can give definite classical trajectories of a particle in the
classical case. In this sense the mean-square deviation of the observable is
zero; and accordingly the expectation value of the position operator defines
a classical path. In next section we will discuss a similar problem in
greater detail in association with the random phase induced by the
Heisenberg's momentum-coordinate uncertainty.

\section{Classicality of Apparatus Due to Position-Momentum Uncertainty}

In the above discussion the quantum realization of measuring process boils
down to the appearance of decoherence in the entanglement between the
measured system and measuring apparatus only due to the coupling of the
measured system with the inner environment rather than that between the
pointer of apparatus and the inner environment.

It is well-known that quantum decoherence can be explained in two ways: the
usual explanation for decoherence in a which-way experiment based on
Heisenberg's position-momentum uncertainty relation[15], and the current
explanation based on quantum entanglement, which is not related to this
uncertainty relation directly. In the latter explanation the quantum
correlations between the environment and the considered system are
responsible for the destruction of quantum coherence. In the present paper
the considered system is a composition formed by the measured system plus
the pointer. In fact, in the previous sections we have adopted the second
viewpoint to deal with quantum measurement problem. In this section we will
see that the first explanation can also work well in our modelled quantum
measurement problem. To this end , we first show that the back action of the
inner environment on the system plus pointer implied by Heisenberg's
position-momentum uncertainty relation will disturb the phases of states of
the system plus pointer.This result [16]

Let us return to our model mentioned in the last section. We assume that the
initial state $\mid j\rangle $ of each component of the inner environment is
a real wave packet, which is symmetric with respect to both the canonical
coordinate $\eta _j$ and the corresponding canonical momentum $p_j.$
\begin{equation}
\langle \eta _j\rangle \equiv \langle j|\eta _j\mid j\rangle =0,\quad
\langle p_j\rangle =0
\end{equation}
We will not need its concrete form. Rather we assume it to be of Gaussian
type with the variance $a_j$ $=\Delta \eta _j$ in $\eta _j-space.$ Here, we
adopt the definition of the standard deviation
\begin{equation}
\Delta \phi \equiv \sqrt{\langle (\phi -\langle \phi \rangle )^2\rangle }=%
\sqrt{\langle \phi ^2\rangle -\langle \phi \rangle ^2}
\end{equation}
for a given phase $\phi .$ Physically, once $\Delta \eta _j$ is given , the
variance of $p_j$ cannot be arbitrary since there is Heisenberg's
position-momentum uncertainty relation $\Delta \eta _j\Delta p_j\geq \frac
12.$In the last two sections, the conclusion drawn seems to depend on the
choice of the concrete form of the initial state, but now we can argue that
this is not the case with the above consideration.

After a measurement, the final state of the $j^{\prime }s$ environment
component $\eta _j$ entangling with the system-pointer state $|n\rangle
\otimes $ $\left| x_n\right\rangle $ is just
\begin{equation}
\left| e_n^{[j]}\right\rangle =U_{en}^j(t)|j\rangle .
\end{equation}
The $j\prime $th factor in the decoherence factor $F_{mn}=\langle
e_n|e_m\rangle $ =$\prod_{j=1}^NF_{mn}^j$ can be written as
\begin{equation}
F_{mn}^j\equiv \langle e_n^{[j]}|e_m^{[j]}\rangle =\langle j|\exp [i\hat{\phi%
}_{mn}^{[j]}]|j\rangle
\end{equation}
by the Wei-Norman algebraic method. Here, the time dependent
global phase is neglected. We can understand the Hermitian
operator
\begin{equation}
\hat{\phi}_{mn}^{[j]}=(n-m)f_it\hat{D}_j
\end{equation}
in terms of the generalized phase difference between $\left|
e_m^{[j]}\right\rangle $ and $\left| e_n^{[j]}\right\rangle $ for
\begin{equation}
\hat{D}_j=\frac t{2m_j^{^{\prime }}}\hat{p}_j+\eta _j.
\end{equation}
The standard deviation
\begin{equation}
\Delta \hat{\phi}_{mn}^{[j]}=(n-m)f_jt\Delta D_j
\end{equation}
is proportional to time $t$ and represents a random phase change of $%
|n\rangle \otimes $ $\left| x_n\right\rangle $ by the $j^{\prime }th$ inner
component.

The whole random phase change of $|n\rangle \otimes $ $\left|
x_n\right\rangle $, contributed by the inner variables, is determined by

\begin{equation}
\hat{\phi}_{mn}=\sum_{j=1}^{N-1}\hat{\phi}_{mn}^{[j]}
\end{equation}
Physically, each variable of the inner environment can exert a different
impact independently on the different components of the entangling states of
the measured system plus the pointer. If we think each uncertain phase
change by this perturbation as an independent stochastic variable, we have
\begin{equation}
\Delta \hat{\phi}_{mn}=\sqrt{\sum_{j=1}^{N-1}(\Delta \hat{\phi}_{mn}^{[j]})^2%
}\geq \sqrt{N}\min \{(\Delta \hat{\phi}_{mn}^{[j]})^2|j=1,2..N-1\}
\end{equation}
We observe that the phase uncertainty $\Delta \hat{\phi}_{mn}$ caused by all
inner variables can be amplified to a number much greater than $2\pi $ when $%
N\rightarrow \infty .$ Then the system-pointer states acquire a very large
random phase factor. Therefore, the inner environment washes out the
interference of any two components of the system plus the pointer .

In the above reasoning we make the connivance that there exists a finite
minimum of $\Delta \hat{\phi}_{mn}^{[j]}$ . This point is just guaranteed by
Heisenberg's position-momentum uncertainty relation $\Delta \eta _j\Delta
p_j\geq \frac 12.$ In fact, because $D_j$ is a linear combination of $\eta
_j $ and $p_j$ and
\begin{equation}
\langle p_i\eta _i\rangle +\langle \eta _ip_i\rangle =0
\end{equation}
for the real wave function average, there should be a quantum limit for its
variance $\Delta D_j:$%
\begin{eqnarray*}
\Delta D_j. &=&\sqrt{(\Delta \eta _j.)^2+(\frac t{2M}\Delta \hat{p}_j)^2} \\
&\geq &\sqrt{(\Delta \eta _j.)^2+\frac 1{(\Delta \eta _j.)^2}(\frac t{2M})^2}
\end{eqnarray*}
If one wishes the highest possible for $D_j,$ one should not make $(\Delta
\eta _j.)^2$ arbitrarily small, because this will make $\Delta p_j$
arbitrarily large . So it is optimal to take $(\Delta \eta _j.)^2=\frac
t{2M} $, and we have the finite minimum $\sqrt{\frac tM}.$ So we have a
minimum of phase uncertainty
\begin{equation*}
\Delta \hat{\phi}_{mn}=\sqrt{N}\frac tM
\end{equation*}
This result qualitatively illustrates the many-particle amplification effect
of uncertain phase change.

The direct relationship between the two explanations for decoherence can
also be revealed explicitly in our present model. As a matter of fact, this
problem has been tackled by Stern, Aharonov and Imry,with the observation$%
\langle e_n|e_m\rangle =\langle e^{i\hat{\phi}_{mn}^{[j]}}\rangle $ [17].
Consider the specially -chosen initial state of Gaussian type $\langle \eta
_j\mid j\rangle =\left( \frac 1{2\pi a^2}\right) ^{\frac 14}\exp (-\frac{%
\eta _j^2}{4a^2}).$ Since the standard deviation $\Delta \eta _j$ is the
width $a$ of Gaussian wave packet and the uncertainty $\Delta p_j$ of the
momentum fluctuation is $\frac 1{2a}$ , we have
\begin{equation*}
\left( \Delta D_i\right) ^2=\left( \Delta \frac t{2m_i}p_i\right) ^2+\left(
\Delta \eta _i\right) ^2=\left( \Delta D_i\right) ^2=\frac{t^2}{16m_i^2a^2}%
+a^2
\end{equation*}
Then we can probe the relationship between the two explanations by using the
exact solution in the last section
\begin{eqnarray}
F_{mn} &=&\prod_{j=1}^{N-1}\exp [-(n-m)^2f_j^2t^2\frac 18(\frac 1{(\Delta
p_j)^2}+\frac 1{(\Delta \frac{2m_i^{^{\prime }}}t\eta _j)^2})]  \notag \\
&=&\prod_{j=1}^{N-1}\exp [-\left( n-m\right) ^2f_j^2t^2\frac 12\left( \Delta
D_j\right) ^2]  \notag \\
&=&\exp \left[ -\frac 12\sum_{j=1}^{N-1}\left( \Delta \phi _{mn}^j\right) ^2%
\right] =\exp \left[ -\frac 12\left( \Delta \phi _{mn}\right) ^2\right]
\end{eqnarray}
This result just shows that the condition that $F_{mn}=0$, which is required
by an ideal measurement implies a large random phase variance $\left( \Delta
\phi _{mn}\right) ^2.$ Furthermore, we can conclude from this exact solution
that the large random phase change just originates from Heisenberg's
position-momentum uncertainty relation $\Delta \eta _j\Delta p_j=\frac 12.$
In the terminology of classical stochastic process, $\eta _j$ and $p_j$ can
be regarded as a pair of uncorrelated stochastic variables, but the
uncertainty relation $\Delta \eta _j\Delta p_j=\frac 12$ exerts a constraint
on them. This constraint just reflect the uncertainty of phase change in the
measuring process.

\section{Concluding Remarks}

The above arguments shed a new light on the understanding of the
relationship between Bohr's complementarity principle and Heisenberg's
uncertainty principle. It is well known that the principle of
complementarity usually refers to the wave-particle duality in quantum
mechanics. It says that the quantum-mechanical entity can behave as particle
or wave under different experimental conditions,but these two natures
excluded each other in the experiment. For example, in the famous Yang's
double-slit experiment, the matter wave of a single particle can apparently
pass through both slits simultaneously. In this sense the experiment
emphasizes the nature of wave and so there \ forms an interference pattern.
On the contrary, if a `which-way'detector is employed to determine the
particle's path, the interference pattern is destroyed. \ This is because
''which-way' experiment focuses on the nature of particle with classical
location,''path''. This can further be explained in terms of Heisenberg's
uncertainty principle: the uncertainty in particle's momentum will introduce
random phase difference between two paths and thus destroys the
interference. At this point it is worth mentioning that Durt et al[18]
reported a which-way experiment in an atom interferometer in which the `back
action' of path detection on the atom's momentum is too small to explain the
disappearance of the interference pattern.

We would also like to emphasize that the arguments about the quantum
description of the macroscopic object such as the moving mirror is closely
related to the Schroedinger$^{^{\prime }}$s cat problem [19,20] , a
conventional topic of quantum decoherence concerning the state superposition
for macroscopic object. Most recent progress has been made in demonstrating
Schrodinger$^{^{\prime }}$s cat state in various macroscopic quantum systems
such as superconductors , laser-cooled trapped ions, photons in a microwave
cavity and $C_{60}$ molecules[21]. According to the viewpoint in this paper,
a system with many microscopic degrees of freedom can behave quantum
mechanically only if it is sufficiently decoupled from the environment and
the phases of its inner states match very well.

In sum, we have proposed an alternative Zurek's triple approach for quantum
measurement with time independent Hamiltonian. The calculation for a
specific model shows the possibility of implementing our approach. It should
be pointed out that in the above discussed extremely-idealized model the
interaction between the particles composing the moving mirror has not been
considered. But we don't think this is a big problem. Starting with the main
idea developed in this paper, we can discuss more general situation with
inter-particle interaction in a similar way. Although the exactly solvable
model in this paper might not exist in reality, the idea of quantum
adiabatic separation or the master equation is physically meaningful.

\bigskip

\textbf{Acknowledgement:} \textit{This work is supported by the NSF of China
and the knowledged Innovation Programme(KIP) of the Chinese Academy of
Science.}

\end{document}